\documentclass[aps,prx,reprint,superscriptaddress]{revtex4-2}

\usepackage{silence}
\usepackage[dvipsnames]{xcolor}
\usepackage{float}
\usepackage{graphicx}
\usepackage{color}
\usepackage{amssymb}
\usepackage{amsmath}
\usepackage{mathtools}
\usepackage{accents}
\usepackage{hyperref}
\usepackage{lipsum}
\usepackage{subfiles}
\usepackage{physics}
\usepackage{tikz}
\usepackage[normalem]{ulem}
\usepackage{pgfplots}
\usepgfplotslibrary{statistics}
\usepgfplotslibrary{groupplots}
\usetikzlibrary{plotmarks}
\usetikzlibrary{arrows.meta}
\usepackage{pgfplotstable}
\usepackage{verbatim}
\usepackage{xr}

\graphicspath{{img/}}

\newcommand*{\kbt}{k_B T}

\newcommand*{\median}[1]{\mathrm{Md}\left(#1\right)}

\newcommand*{\kloop}{k_l}
\newcommand*{\koff}{k_o}
\newcommand*{\kbind}{k_b}
\newcommand*{\kunbind}{k_u}

\newcommand*{\Eass}{E_a(j)}
\newcommand*{\Ebound}{E_b(j)}
\newcommand*{\dGb}{\Delta G_b(j)}

\newcommand*{\kjump}{\omega(i,j)}
\newcommand*{\kreset}{r(j|i)}

\newcommand*{\dGl}{\Delta G_l}

\tikzset{>=latex}

\pgfplotsset{compat=newest}

\begin{document}
\title{A general mechanism for enhancer-insulator pairing reveals heterogeneous dynamics in long-distant 3D gene regulation
}
\date{\today}

\author{Lucas Hedstr\"{o}m}
\email{lucas.hedstrom@umu.se}
\affiliation{Integrated Science Lab, Department of Physics, Ume\r{a} University, SE-901 87 Ume\r{a}, Sweden}
\author{Ralf Metzler}
\email{rmetzler@uni-potsdam.de}
\affiliation{Institute for Physics \& Astronomy, University of Potsdam, DE-144 76 Potsdam-Golm, Germany}
\author{Ludvig Lizana}
\email{ludvig.lizana@umu.se}
\affiliation{Integrated Science Lab, Department of Physics, Ume\r{a} University, SE-901 87 Ume\r{a}, Sweden}

% ==========================================================================
% Abstract
\begin{abstract}
	Cells regulate fates and complex body plans using spatiotemporal signaling cascades that alter gene expression. Enhancers, short DNA sequences (50-150 base pairs), help coordinate these cascades by attracting regulatory proteins to enhance the transcription of distal genes by binding to promoters. In humans, there are hundreds of thousands of enhancers dispersed across the genome, which poses a challenging coordination task to prevent unintended gene activation. To mitigate this problem, the genome contains additional DNA elements, insulators, that block enhancer-promoter interactions. However, there is an open problem with how the insulation works, especially as enhancer-insulator pairs may be separated by millions of base pairs. Based on recent empirical data from Hi-C experiments, this paper proposes a new mechanism that challenges the common paradigm that rests on specific insulator-insulator interactions. Instead,  this paper introduces a stochastic looping model where enhancers bind weakly to surrounding chromatin. After calibrating the model to experimental data, we use simulations to study the broad distribution of hitting times between an enhancer and a promoter when there are blocking insulators. In some cases, there is a large difference between average and most probable hitting times, making it difficult to assign a typical time scale, hinting at highly defocused regulation times. We also map our computational model onto a resetting problem that allows us to derive several analytical results. Besides offering new insights into enhancer-insulator interactions, our paper advances the understanding of gene regulatory networks and causal connections between genome folding and gene activation.
\end{abstract}

\maketitle

%\tableofcontents

% ==========================================================================
% Introduction
\section{Introduction}
\begin{figure*}
    \centering
    \includegraphics[width=0.8\textwidth]{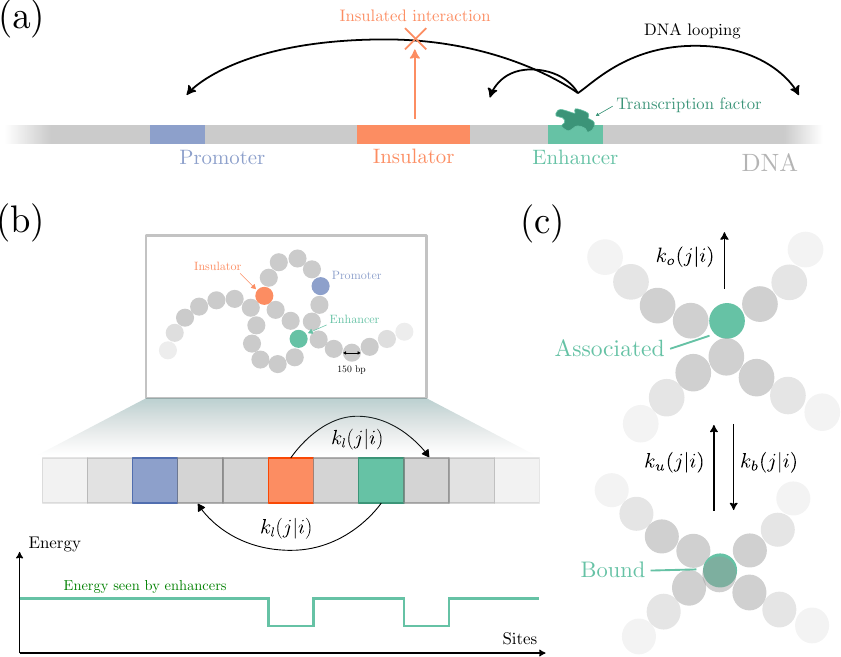}
    \caption{Schematics of enhancer-promoter-insulator constructs on DNA (a) and our three-state looping model (b)-(c). 
    \textbf{(a)} The insulator region blocks long-ranged enhancer-promoter interactions. Reducing such interactions causes "insulation" as the transcription factor cannot physically assist the transcription machinery (e.g., RNA polymerases) that interacts with the promoter.
    \textbf{(b)} Three-state lattice model.  We model the chromatin as a sequence of $N$ sites.  A few of these sites represent enhancers (green) or insulators (orange). They can loop and associate to any other site with the rate $\kloop(j|i)$. This rate balances looping entropy and binding energy,  where this energy changes with their time-varying relative positions. The line graph below shows a snapshot of the energy landscape experienced by the enhancer.  This landscape has two "dips" corresponding to the insulator loop anchor.
    The other sites represent "inactive" chromatin. Unlike enhancers and insulators,  we consider chromatin as an ensemble of interacting loops and include them by adjusting the enhancer-promoter looping exponent  \cite{schafer1992renormalization,hanke2003entropy}.  We color the promoter in blue.  In the model, we treat the promoter as an absorbing target. 
    \textbf{(c)} Local binding dynamics. After looping, the enhancer and insulator can either strongly associate (bind) to the new site by $\kbind$ or dissociate back to its original position by $\koff$ (similar to a resetting process).  If bound, the return rate to the associated state is $\kunbind$. 
    }
    \label{fig:model}
\end{figure*}

Cell fates and complex body plans are established through signaling cascades that turn genes on and off in complex spatiotemporal patterns.  One of the critical genetic elements that help coordinate these cascades is enhancers.  These are short regulatory DNA sequences (50--150 base pairs (bp)) that attract proteins, such as transcription factors, to "enhance" the transcription of select genes (Fig. \ref{fig:model}(a)). Enhancer elements are often far from the target gene start, sometimes as far as millions of base pairs apart.  Yet, experiments show they appear close in 3D to regulate transcription \cite{chakraborty2023rewiring}.

In humans,  the genome harbors hundreds of thousands of dispersed enhancers supporting gene expression networks \cite{panigrahi2021mechanisms}, Notably, these enhancers do not necessarily act on the closest promoter and may regulate multiple genes \cite{geyer2002protecting, mohrs2001deletion}.  This posits a challenging coordination task of all these distal 3D interactions to protect genes from unintended activation. One way cells manage this task is by using insulators. Like enhancers,  insulators contain clusters of binding sites for sequence-specific DNA-binding proteins that block enhancer-promoter interactions. However,  insulators are typically a bit larger, spanning approximately $\sim 300-2000$~bp.  When first discovered in \textit{Drosophila melanogaster}~\cite{udvardy198587a7,  holdridge1991repression, bushey2009three}, the insulators appeared to define boundaries between different chromatin states. However, researchers soon found they could block enhancer action when inserted at specific genomic loci and that gene activity depended on specific DNA binding proteins associating with the insulator element.  

From a genetic point of view,  insulators are simply some DNA piece that activates a gene when removed. But the question is how this insulation works mechanistically, especially as some enhancer-promoter distances are so large \cite{chakraborty2023rewiring}. One of the most popular mechanistic descriptions is the topological model. This model suggests that two or more insulating elements bind each other to form loops \cite{raab2010insulators}. This idea agrees with Hi-C data from mammals, where CTCF insulator elements (CCCTC-binding factor) make 3D contacts and often define borders of shielded chromatin communities, so-called Topologically Associated Domains \cite{rao20143d, dixon2012topological}.  This is further consistent with the loop-extrusion model, where a hand-cuff-shaped protein (Cohesin) binds and extrudes DNA through itself until it reaches a CTCF site, thus creating a loop with CTCF as the anchor points \cite{sanborn2015chromatin,polovnikov2023topological}. The topological model also agrees with extensive polymer simulation studying insulation in varying enhancer-insulator-promoter configurations (measured by reduced 3D contact probabilities) \cite{doyle2014chromatin}.

However, there are recent experimental data that challenge this paradigm \cite{kahn2023topological}. Using Hi-C experiments from \textit{Drosophila melanogaster} to measure contacts across thousands of enhancer-insulator pairs, the paper convincingly shows that the data is inconsistent with specific looping interactions between insulators (at least, much higher than with general chromatin). This suggests the topological insulator model needs revision (at least in \textit{Drosophila melanogaster}).  Building on this observation,  our paper proposes an alternative mechanism where insulators bind weakly to surrounding chromatin rather than other insulators.  We formulate our model on a lattice with stochastic looping dynamics where we calibrate the rates to existing experimental data and benchmark to existing measured contacts from \cite{kahn2023topological}. Next,  we use our model to study the dynamics of enhancer-promoter hitting times and show that the average may deviate substantially from the typical (most probable). We also map our computational model onto a resetting problem and provide several analytical results. Our work offers new insights into the enhancer-insulator mechanics. Besides yielding a better understanding of gene regulatory networks,   knowing how insulators work may help unveil causal relationships connecting gene expression and genome folding \cite{schwartz2017three, dekker20163d}.

% ==========================================================================
% Methods
\section{Methods}

% =============================================================
\subsection{Looping model for enhancer-insulator dynamics}\label{sec:model}

    We represent the chromatin as an array of sites $i = 1,\ldots, N$, where each site symbolizes a nucleosome ($\approx$ 175 bp), the basic chromatin unit.  This choice is natural from a biological (or epigenetic) point of view \cite{lizana2023polycomb, lundkvist2023forecasting, berry2017slow} but not critical for our general framework.  The array has four site types: enhancers, promoters,  insulators, and regular chromatin (Fig.~\ref{fig:model}).  In most simulations, we consider one enhancer and one promoter, typically placed 20-50 array indices between one another ($\approx 3.5-8.5 \times 10^3$~kb), and up to twenty insulators. However, actual gene clusters usually have a much richer arrangement where several enhancers and insulators act in concert to ensure proper gene expression of many genes. But to better appreciate the model, we study simpler configurations.

    As mentioned above, enhancers are DNA elements that attract regulatory proteins, such as transcription factors. While being attached to DNA,  the transcription factors try to find the promoters associated with the designated promoter to regulate transcription.  In other words,  protein-bound enhancers make repeated looping attempts with surrounding chromatin until they reach the target site.

    Our model treats the promoter as an absorbing point, and the simulation stops once the enhancer complex reaches there (i.e., infinite reaction rate).  This contrasts the interaction with surrounding chromatin which is much weaker.  We model enhancer-chromatin binding using standard transition-state theory where the enhancer may be bound with energy $E_b$ or loosely associated ($E_a$) from where it may detach (Fig. \ref{fig:model}). These two states ("bound" and "associated") are similar to "search" and "recognition" modes often used to represent two-state searchers to explain the so-called speed-stability paradox in DNA-target search problems \cite{benichou2011intermittent}. In addition to $E_a$ and $E_b$,  there is an energy barrier $\Delta G_b$ separating the bound and associated states (called activation energy in transition-state theory).  Below (Sec. \ref{sec:matching-params}), we estimate these parameters from actual transcription factor binding data.

    In addition to surrounding chromatin, the enhancer also binds to insulators. We model this binding using transition state theory but assign different values for $E_a$, $E_b$, and $\Delta G_b$.  Importantly, because the insulators are dynamic objects like the enhancer, they form loops with surrounding chromatin (discussed below), meaning these energies change with site index $j$ (and time).  Therefore, for some fixed insulator configurations, we express the unbinding and binding rates as
    \begin{equation}\label{eq:binding-rates}
        \begin{split}
        \kunbind(j) &= \gamma e^{-(\dGb - \Ebound)}, \\
        \kbind(j) &= \gamma e^{-(\dGb - \Eass)},
        \end{split}
    \end{equation}
    where $\gamma$ is the basal binding rate, on the order $10^7$ s$^{-1}$\cite{cencini2017,kalodimos2002residue},  and the thermal energy $\kbt$ is set to unity.  Since  $\kbind(j)/\kunbind(j) = \exp(\Eass-\Ebound)$, these rates obey detailed balance.

    Next, we discuss the enhancer's looping rates.  Similar to previous work \cite{benichou_searching_2009,felipe2021dna,shvets2016role,vemulapalli2021dna},  these rates depend on the entropy cost of forming the loop $k_B\ln [(\ell/\ell_0)^{-\alpha}]$ where $\ell$ is the loop length, $\alpha$ is the looping exponent, and $\ell_0$ is a typical loop scale \cite{hanke2003entropy}. 
    To calibrate constants to obey typical DNA-looping times,  we again use transition-state theory and add a small binding activation energy $\dGl$. Given these parameters, the looping on-rate from lattice site $i$ to $j$ is
    \begin{equation} \label{eq:looping-rates}
        \begin{split}
            \kloop(j|i) =
            \begin{cases}
                \delta e^{- \alpha \ln  (\ell_{ij}/ \ell_0) - \dGl} & \text{if } i \neq j\\
                0 & i=j,
            \end{cases}
        \end{split}
    \end{equation}    
    where $\delta$ is the basal looping frequency, which is on the order of $10^3$s$^{-1}$ \cite{cencini2017}, and $\ell_{ij} = d_\textrm{nucl.}\times|i-j|$ is the loop length;  $d_\textrm{nucl.}$ is the nucleosome diameter ($\sim 10$~nm, or 175 bp). 

    The looping off-rate follows a similar formula as Eq. \eqref{eq:binding-rates}, but it contains the energy activation associated with the looped state $\dGl$ instead of $\dGb$ since we imagine this is the only state from which the loop can break apart (it is also unphysical that the loop rate depends on the position-dependent $\Eass$).  Thus, to unloop from the bound state ($\Ebound$), the enhancer must first become "associated" ($\Eass$) and then unloop. In summary, the looping off-rate is
    \begin{equation}\label{eq:off-rates}
        \koff(j|i) = \delta e^{- (\dGl - \Eass)}
    \end{equation}
    where the two looping rates obeys detailed balance  $\kloop(j|i)/\koff(j|i) = \exp(-\Eass - (-\alpha \ln  (\ell_{ij}/ \ell_0)))$.

    Before showing how we calibrate the binding energies to experimental data, we make three comments. First, even if the discussion above was mostly about enhancers,  we assume insulators to follow the same dynamics (Eqs. \eqref{eq:binding-rates}-\eqref{eq:off-rates}), albeit with slightly different energy parameters. For example, we assume that insulators interact weekly with chromatin.  They do so with a binding constant that should not be smaller than $K\sim 10^2$~$\mu$M, which is the typical scale for specific binding (estimated from E. coli \cite{phillips2012physical}).  While our model does not rest on specific insulator-insulator interactions, they are not excluded. But they cannot be significantly larger than for general chromatin. If so,  insulator pairs would appear as high-contact stripes in Hi-C maps, which was not observed in \cite{kahn2023topological}. Therefore, we let these interactions have the same strength as regular insulator-chromatin interactions.

    We also point out that the insulators in our simulations constantly form loops and, therefore, appear on two different lattice positions (Fig. \ref{fig:model}(b)) from the point of view of the enhancer. One position is always fixed and coincides with the insulator's designated DNA segment. The second one represents the other end of the insulator-chromatin loop.  This loop is short-lived, so this second position is highly dynamic and switches frequently and symmetrically around the insulator's primary position during the simulation  (Fig. \ref{fig:ana-vs-sim}). This effect implies that the enhancer has two possibilities to bind the insulator (Fig. \ref{fig:model}(c)).

    Second, thus far, we have discussed DNA looping, omitting regular chromatin. But in reality, chromatin also fluctuates.  Instead of introducing specific looping rates akin to Eqs. \eqref{eq:looping-rates} and \eqref{eq:off-rates}, we treat chromatin as an ensemble of interacting loops and modify the looping exponent $\alpha$ for enhancers and insulators accordingly. For long self-avoiding chains, the exponent (often denoted "ring factor") is
    \begin{equation}
            \alpha \approx d \nu - 2\sigma,
    \end{equation}
    where $d$ is the embedding dimension, $\nu$ is the exponent associated with the polymer's radius of gyration, and $\sigma$ is the "scaling factor" ($\sigma=0$ for non-interacting loops) \cite{kafri2002denaturation}. Using $d=3$, $\nu=0.588$ and $\sigma=-0.175$ gives $\alpha = 2.114$ \cite{schafer1992renormalization}. 

    Third,  some papers include a bending term  $\propto 1/\ell_{ij}$  in the looping rates \cite{felipe2021dna}. This contribution accounts for the bending energy cost when creating short loops.  We omit this since we only consider loops which are typically much longer than the DNA's Kuhn length ($\approx 300$ bps). 

\begin{figure}
    \centering
    \includegraphics[width=0.8\columnwidth]{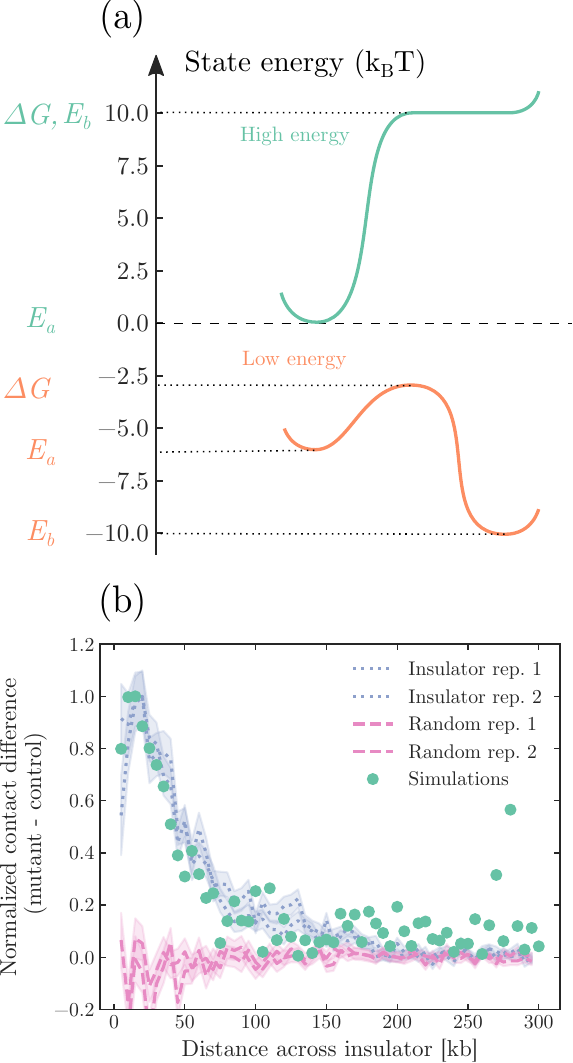}
    \caption{Energy levels and fit to experimental data \cite{kahn2023topological}. 
    \textbf{(a)} The figure shows two typical interaction cases: high and low energy (green and orange)  We found that the interaction energy is high for non-specific sites (grey sites in Fig.~\ref{fig:model}). This corresponds to a weak association energy ($E_a)$ and a large barrier $\Delta G$ separating the bound ($E_b$) and associated state. This case differs from enhancers and insulators, where the interaction energy is low (orange lines). Here, the association and bound energy are both large, separated by a smaller energy barrier.
    \textbf{(b)}. Recent experiments found that the contact difference between DNA regions separated by insulator-rich segments increases if the insulators are compromised (i.e., genetically engineer a mutant lacking the CP190 binding factor that binds to insulators)~\cite{kahn2023topological}. We replot the empirical data here (with permission) as dotted blue and purple lines.  The shaded areas represent $\pm$ one standard deviation.  To match the data, we simulated a similar system with a center-placed insulator region and calculated the probability density functions of each site flanking the insulator. The y-axis is normalized by dividing by the maximum value of our simulations and the experimental data. By considering that each site is on the same scale as the Hi-C data used in experiments (5kb), we found an excellent agreement between our simulations and the data considering a stickiness between the insulator-rich region and other regions ($E_{\rm sticky} = -17$) and fast looping dynamics ($l_0 = 0.4$). %
    }
    \label{fig:energy-and-contacts}
\end{figure}

% ==========================================================================
% Results
\section{Results}
% =================================
\subsection{Matching model parameters to  data}\label{sec:matching-params}

    To calibrate the model to empirical data, we estimate $\Eass$, $\Ebound$, and $\dGb$ for binding and association, and $l_0$ and $\dGl$ for the looping process. To this end, we use comprehensive binding data for transcription factors and measured \textit{in vitro} looping rates.  We start with association and binding.

    Previous work calculated the energy (and free-energy) landscape along DNA for individual transcription factors (TFs) that bind to sequence-specific motifs, typically  10-30 bp long \cite{cencini2017,stormo1998specificity,berg1987selection}. But because DNA is so much longer (and there are only four base pair types), there are many instances of almost similar binding sequences. This results in fluctuating genome-wide binding profiles interpreted as the TF's spatial probability density $p_\textrm{TF}(x)$, where $x$ denotes the DNA coordinate. From this probability density, it is common to define the energy profile ${\cal E}(x)$,  assuming that  $p_\textrm{TF}(x)\sim \exp(-{\cal E}(x))$. We defer to \ref{supp-sec:typical-energies} and \cite{cencini2017} for technical details about obtaining  ${\cal E}(x)$ from a given TF target sequence. In \ref{supp-sec:typical-energies}, we plot the average binding data for about 300 TFs that we use to estimate $\Eass$, $\Ebound$, and $\dGb$ for the enhancers and insulators. In our lattice notation ${\cal E}(x) \to E(j)$.

    To estimate $\Eass$, $\Ebound$, and $\dGb$ for the enhancer, we use the formalism developed \cite{cencini2017} for a two-state TF flipping between "search" and "recognition" mode while searching for a DNA-target sequence.  When in recognition mode, the TF is immobile, and the residence time depends on how similar the local and target sequences are to each other.  In other words, the time depends on the depth of the landscape $E(j)$.   When in search mode, the TF diffuses and only weakly interacts with the DNA.  In \cite{cencini2017}, it is assumed that the effective energy landscape while diffusing is a scaled version of $E(j)$, i.e., $\rho E(j)$, where $\rho$ is adjusted to agree with measured 1D diffusion constants.  In practice, this means setting $\rho \leq 0.3$; we use $\rho = 0.3$.  Lastly, there is a free energy barrier $\Delta G_{\mathrm{RS}}$ separating "search" and "recognition" mode that we also extract from TF-binding data (see below and in \ref{supp-sec:typical-energies}). In summary,  the  equations to calculate $\Eass$, $\Ebound$, and $\dGb$ read
    \begin{equation}\label{eq:energies}
        \begin{split}
            &\Eass = \rho E(j), \\
            &\Ebound = \Delta G_{\mathrm{RS}} + E(j), \\
            &\dGb = \Delta G_{\mathrm{RS}} + \frac{1+\rho}{2} E(j),
        \end{split}
    \end{equation}
    where all the parameters on the right-hand side come from empirical data. In particular,  we used TF data from the JASPAR database \cite{fornes2020jaspar} to extract  $E(j)$ and $\Delta G_{\rm RS}$. We apply these formulas to the three binding instances we have in our problem: enhancer-insulator, insulator-insulator, and unspecific (enhancer-chromatin and insulator-chromatin.) We used $\Delta G_{\rm RS} = 10.13$ in all three cases,

    \textit{Enhancer-insulator binding} We calculated $\langle E(j) \rangle_{\mathrm{binding\ sites}}$ for all human chromosomes and available transcription factors from the JASPAR database \cite{fornes2020jaspar}. Using Eq. \eqref{eq:energies},  we find that the population median ($\median{\cdot}$) is $\median{\langle E(j) \rangle_{\mathrm{binding\ sites}}} = -21.13$ (see \ref{supp-sec:typical-energies}.) This gives $\Eass = -6.34$, $\Ebound = -11.0$, and $\dGb = -3.60$, which is a system having low binding energy, corresponding to a low association energy and energy barrier (see Fig~\ref{fig:energy-and-contacts}).

    \textit{Unspecific binding for enhancers}. Here we set $E(j) = 0$. Plugging this into Eq.~\eqref{eq:energies} gives $\Eass = 0$, $\Ebound = 10.13$ and $\dGb = 10.13$. This situation is where strong binding is rare and weak; see Fig~\ref{fig:energy-and-contacts}. 

    \textit{Unspecific binding for insulators}. Here we set $E(j) = -15.0$. Plugging this into Eq.~\eqref{eq:energies} gives $\Eass = -4.5$, $\Ebound = -4.9$ and $\dGb = 0.39$.

    Next, we match the looping rates for enhancers and insulators. To set $l_0$ and  $\dGl$, we use experimental data from \cite{finzi1995measurement,broek2006real,weintraub2017yy1,dekker20163d}, reporting that typical looping times for ~300 bps long loops are $10^1-10^2$ seconds (LacI protein and restriction enzymes NaeI and NarI), and $10^3$ seconds for ~3500 bp long loops. Matching with our simulation gives $l_0 \approx 0.04$ and $\dGl=0.0$. However, the looping times \textit{in vitro} may deviate substantially in crowded cell conditions \cite{dekker20163d}, where far-away sections come in contact faster than expected. To this end, we vary $l_0$ to study the fast- and slow-looping regimes.

% =================================
\subsection{Sticky insulators reproduce measured contact differences in \textit{Drosophila melanogaster} embryos}

    In this section, we benchmark our model to empirical data from \cite{kahn2023topological}. Using Hi-C experiments, this paper quantifies how effectively insulators block 3D interactions between flanking DNA segment pairs. By collecting contact profiles for hundreds of insulator positions in mutant and wild-type Fruit fly embryos, the authors made two key observations. First, the insulators block 3D interactions over distances up to 200 kb. We replot the empirical data in Fig.~\ref{fig:energy-and-contacts}(b) for two replicate experiments (Rep. 1 and Rep. 2).  The background is derived from the same experiments but uses random loci far from any insulators. Second, they also measured 3D contacts between insulator pairs and could not detect any specific binding, which contradicts the standard topological model. 

    To reproduce the measured contact decay in Fig.~\ref{fig:energy-and-contacts}(b) using our model, we constructed a large-scale version where each lattice site matches the resolution of the data (i.e., $5$~kb), rather than a single nucleosome ($\sim 0.2$~kb). In the middle of the lattice, we placed an insulator-rich region (one site) that weakly associates with all other sites with the same energy ($E(j)=\rm {const.}$) and forms loops. To get the contact frequencies, we simulate repeated looping events of all flanking sites across the insulator according to rates $\kloop(i|j)$ and $\koff$ (Eqs. \eqref{eq:looping-rates} and \eqref{eq:off-rates}) and record the residence times between all site pairs. Next, we collected these residence times over $10^2$ simulations ($\approx 10^6$ time steps each) and used them as a proxy for contact probabilities. To fit the model to the dotted lines in Fig.~\ref{fig:energy-and-contacts}(b), we calculated the relative difference in these probabilities with and without the insulator (Fig. ~\ref{fig:energy-and-contacts}(b), black). 

    We note that our model agrees well with the empirical data using $E(j) = -17$ as unspecific binding for insulators and $l_0 = 0.4 \leftrightarrow 2000$bp.  This length scale agrees with typical coarse-grained "beads-on-a-string" polymer models for chromatin. For example, if modeled as a freely jointed chain, each monomer should contain 2000-25000 bp of DNA \cite{mirny2011fractal}.  This value differs slightly from our previous fitting using \textit{in vitro} looping times, where $l_0=0.2\leftrightarrow 1000$ bp. In the remaining part of the paper,   we return to a nucleosome-centric model and use $l_0=0.24$ and the unspecific binding $E(j) = -15$ as the standard settings (albeit we also study the effects of $l_0$ variations). But, in summary, this section shows that our model can reproduce empirical data from Hi-C measurements across an ensemble of insulators in \textit{Drosophila melanogaster}.

    \begin{figure*}[!htbp]
        \centering
        \includegraphics[width=\textwidth]{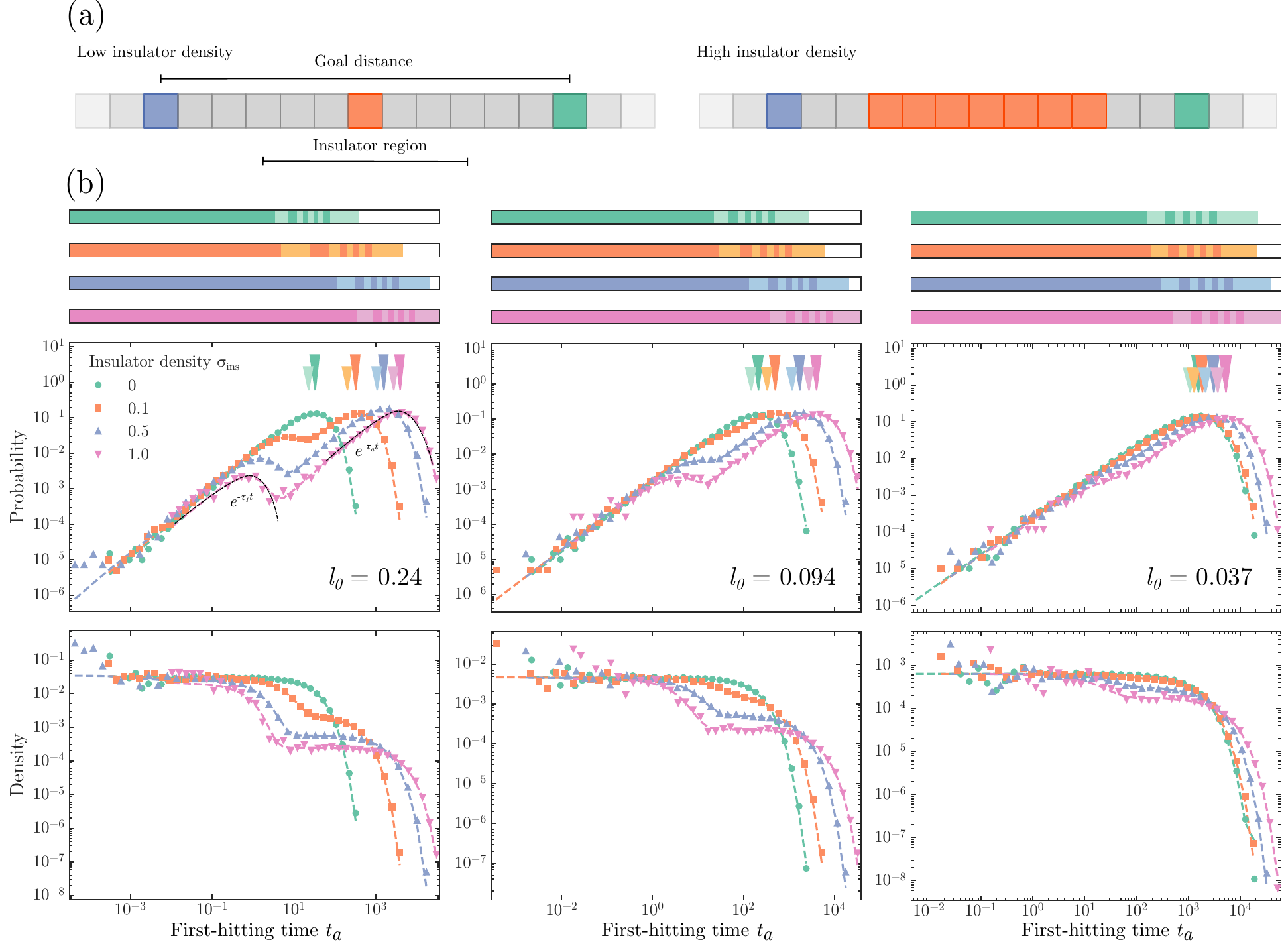}
        \caption{Histograms of the first-hitting times for varying insulator density $\sigma_\textrm{ins}$ and looping scale $l_0$. %
        \textbf{(a)} Schematics of enhancer-insulator promoter configurations for two insulator densities ("low" and "high").  The distance between the enhancer (left, blue) and the promoter (green, right) is 30 sites. The insulators are placed in the middle (orange), where we increase the density by extending the insulator region (from 0 to 20). We used a spacing of 5 sites between the insulators and the enhancer and promoter to avoid effects simply due to proximity. %
        \textbf{(b)} Log-binned histograms of the first-hitting times for different insulator densities. We portray the histograms in two ways: area-normalized (bottom, "Density") and hitting-time probability (see Eq. \ref{eq:hist-prob}) (top, "Probability").  The colors and markers correspond to a varying amount of insulators (a). The colored bars above show the 10th percentiles of all data points.  The filled and shaded arrows above the lines indicate the mean (filled) and median (shaded) first-hitting times, respectively.  We observe a few peaks and valleys in the histograms, corresponding to different search trajectories, such as the enhancer finding the promoter on the first try (potentially with a few intermediate fleeting chromatin interactions) or getting sequestered by the insulator for a significant period. 
        }
        \label{fig:histograms-ins-density}
    \end{figure*}

% =============================================================
\subsection{Insulator densities strongly affect enhancer-promoter hitting frequencies}

    One of the paper's key aims is to better understand the hitting dynamics between the enhancer and promoter elements under insulation, as such an encounter is the primary step in transcription. In particular, we wish to calculate the distribution of time interval lengths between hitting events and study how they change with key variables such as insulator densities, positions and binding strengths. To this end, we perform Gillespie simulations using the rates outlined in Sec. \ref{sec:model}.  The simulations use a 200 lattice site system ($\approx 35$~kbp), where the enhancer and promoter reside in the middle, separated by 30  sites in the standard setting ($\approx 5$~kbp) (Fig. \ref{fig:histograms-ins-density}(a)). On these 30 sites, we put an insulator region with length $n_\textrm{ins}$. A typical simulation produces $\approx 10^5$ samples, where we record the time $t_a$ for the enhancer to reach the promoter site for the first time. 

    We show several simulated histograms in Fig.~\ref{fig:histograms-ins-density}(b) with varying insulator density ($\sigma_\textrm{ins} = n_\textrm{ins}/20 = 0.0, 0.1, 0.5, 1.0$) (see legend).  We portray the histogram in two ways to highlight different features. The lower panels show histograms $\rho(t_a)$, where the area is normalized to unity ("Density").  The upper panels ("Probability") show the enhancer's probability ${\cal P}(t_a)$ of finding the promoter within a specific time interval $\Delta t$. We calculate this probability from the density $\rho $ as
    \begin{equation}\label{eq:hist-prob}
    {\cal P}(t_a) = \int_{t_a}^{t_a+\Delta t}  \rho(u)du \approx \rho(t_a)\times \Delta t,
    \end{equation}
    Above the histograms, we also show the accumulated probability as colored stripes. These stripes provide an intuition for the weight of data points, where each shift into a shaded color indicates 10 percentiles of the data. Lastly, we indicate the average and median hitting times by filled arrows above the ${\cal P}(t_a)$ curves, where the larger arrow indicates the mean.
    
    Consider the leftmost histogram column. Plotted in logscale, we note that the hitting-time distributions are broad and that most data points follow the same trend until $t_a \sim 10^{-1}$ and then spread apart. This means the short-time dynamics are insulator-independent. We interpret this as the enhancer loops into the target after a few short-lived encounters with surrounding chromatin without touching the insulator.  

    Plotting  ${\cal P}(t_a)$ instead of $\rho(t_a)$ unveils a series of peaks corresponding to different types of search trajectories. The left peaks are associated with enhancer-promoter contact events not involving insulators (as explained above).  This contrasts with the right peaks, representing the most probable hitting time. These peaks shift towards larger $t_a$ with growing insulator densities,  making the distributions orders of magnitude broader. 

    Apart from simulated data (symbols), the plots contain several dashed lines with identical colors.  These lines closely follow the data points and represent a numerical inverse Laplace transform of a theoretical search-and-resetting model for $\rho(t_a)$ (Eq. \eqref{eq:rhoa-laplace}), that we derive in Sec. \ref{sec:resetting_model}.  In addition, two black dashed lines show local exponential fits for the two peaks.  In particular,  the right black line follows the simple relationship $\rho(t_a)\sim \exp(-t_a/\tau_a)$, where $\tau_a$ is the mean hitting-time derived analytically (Eq. \eqref{eq:first_moment},  Sec. \ref{sec:resetting_model}).  All dashed curves show a good agreement between simulation and the analytic theory.

    Let's consider all the histograms in Fig.~\ref{fig:histograms-ins-density}(b). Each column shows the hitting-time distribution for varying looping scale $l_0$ (values are indicated in the bottom right). While this parameter is often interpreted as the  Kuhn length,  it is also a proxy for chromatin compaction. As discussed in \cite{dekker20163d}, typical looping scales differ significantly for different types of chromatin folding and density (e.g., space-filling versus self-avoiding). By lowering $l_0$ in our model, thus mimicking less dense folding, we note that the left peaks disappear and that the histograms become insensitive to the insulator density. If $l_0$ increases, the peaks become more pronounced, indicating increasing enhancer-insulator interactions due to less time spent looping (and thus allowing for more possible contacts), resulting in greater insulation.

    \begin{figure}%[!htbp]
        \centering
        \includegraphics[width=0.8\columnwidth]{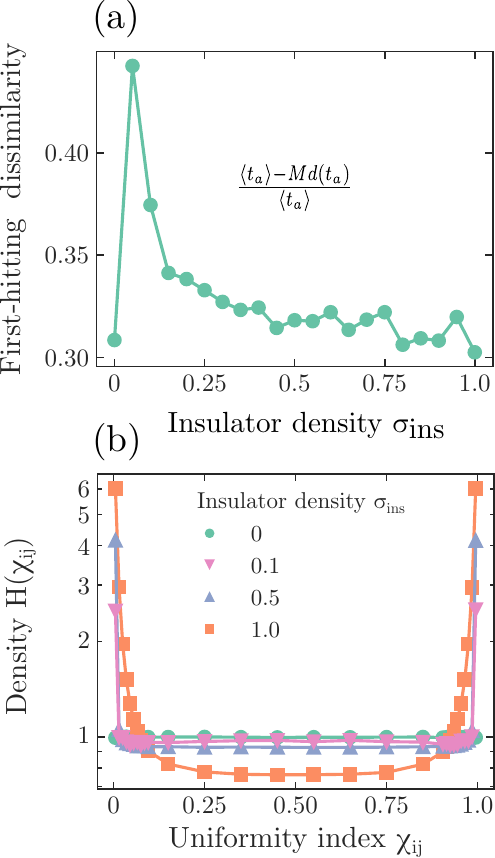}
        \caption{Heterogenous search trajectories. 
        (a) Relative difference between mean and median (see the inset equation) from Fig.~\ref{fig:histograms-ins-density})(b) (filled and shaded arrows). 
        Note the non-monotonic growth and decay. 
        (b) Uniformity index histograms (normalized), $H(\chi_{ij})$ for varying insulator densities.  The bimodal shapes indicate heterogeneous
         enhancer-promoter search trajectories.
         }
        \label{fig:fp-dissimilarity}
    \end{figure}
    
\subsection{Enhancer-promoter search trajectories are highly dissimilar}\label{subsec:dissimilar-search-traj}

    In the previous section, we noted a broad distribution of hitting times $t_a$  (Fig.~\ref{fig:histograms-ins-density})(b)). This suggests that there is a significant difference between average and typical search times (indicated by filled arrows in Fig.~\ref{fig:histograms-ins-density})(b)). To quantify these differences, we plot their relative distance with increasing insulator density $\sigma_\mathrm{ins}$ in Fig.  \ref{fig:fp-dissimilarity}(a). (for fixed $l_0=0.24$). Without insulators, the mean and median are relatively similar (the mean is slightly larger). But as the insulator count increases, the average and the mean start to deviate until reaching some threshold density, where the trend reverses. We interpret this trend change as a result of two competing time scales---one short, governed by chromatin interactions, and one long, dominated by insulator-enhancer interactions.
            
    To further explore the heterogenous difference between search times,  we adopt a theoretical framework based on the uniformity index $\omega_{ij}$ \cite{mattos2014trajectory, grebenkov2018strong, mattos2012first,mejia2011first}. This index is defined as the ratio of one search time versus the sum of two search times for two randomly chosen trajectories $i$ and $j$ 
    \begin{equation}\label{eq:uniformity-index}
        \chi_{ij} = \frac{t_a^i}{t_a^i + t_a^j}
    \end{equation}
    If most samples are similar, the histogram of $\chi_{ij}$ values, $H(\chi_{ij})$, follow a bell-shaped curve centered around $1/2$. However, if the samples are dissimilar, where some are short and others are long,  the distribution is broad or even bimodal with two peaks  $\chi_{ij} = 0$ and $\chi_{ij} = 1$.  We plotted the similarity distribution $H(\chi_{ij})$ in Fig.~\ref{fig:fp-dissimilarity}(b) for several insulator densities.  We see that the histograms are never bell-shaped. Instead, they are bimodal and become increasingly so for growing insulator density ($\sigma_\textrm{ins}$). This shows yet again that there is no typical scale describing enhancer-promoter hitting times. The distribution indicates that there is a significant portion of ultra-short search times where the enhancer immediately loops to the promoter and a large fraction of orders-of-magnitude longer trajectories (note the $t_a$ variation in Fig.~\ref{fig:histograms-ins-density}(b)) where the enhancer gets sequestered at the insulator region, possibly several times, before reaching the promoter site. Because of the heterogeneous dynamics, it is difficult to assign a typical enhancer-promoter hitting time scale describing the breadth of time scales.  Finally,  we point out that this heterogeneity gets reinforced with unsuccessful binding enhancer-promoter binding attempts when they must come together several times before forming a stable complex. However, we do not explore this aspect further here.

% =============================================================
\subsection{Insulation efficiency}

    The histograms in Fig. 2(b) show that the distribution of hitting times changes with insulator density $\sigma_\textrm{ins}$ and looping scale $l_0$. But there is yet another critical variable: the interaction energy $E$ between insulators and enhancers. Here, we explore the relationship between $E$ and the average hitting time $\tau_a$ and where it is most sensitive to changes. 

    To study this question, we calculated $\tau_a$ analytically  (Eq. \eqref{eq:first_moment}) for several values of $E$ and  $\sigma_\mathrm{ins}$, keeping $l_0$ fixed (Fig.~\ref{fig:ana-vs-sim}(a)).  The figure depicts that increasing the insulator density generally leads to higher $\tau_a$. However, the slope is much higher for stronger enhancer-insulator interaction (large negative $E$ values). This suggests that the insulator density has little influence on $\tau_a$ when the interactions are too weak.

\begin{figure}[!htbp]
        \centering
        \includegraphics[width=0.8\columnwidth]{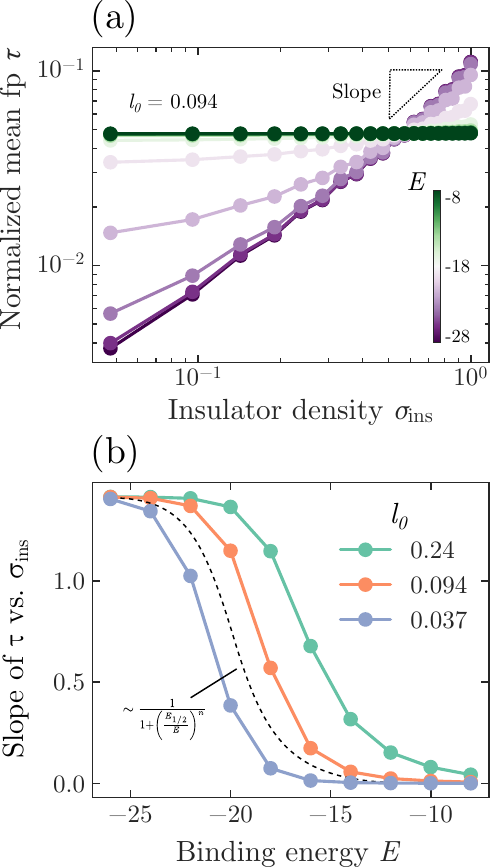}
        \caption{Insulation efficiency
        (a) Average enhancer-promoter hitting time for varying enhancer-insulator interaction energies ($E$, see color bar) and insulator density. Note how the slope increases with strong interactions (smaller $E$).
        (b) Slope sensitivity of $\tau_a$ to $E$, calculated from the large  $\sigma_{ins}$ behavior (see (a)) For all three $l_0$ values, the curves show the same sigmodal behavior with a strong $E$. The dashed line represents an inverted Hill curve, which after fitting gives $E_{1/2} = -15.8,\ -18.3,\ -21.0$ and $n = 10.7,\ 16.3,\ 20.6$ for the three $l_0 = 0.24,\ 0.094, 0.037$ values.
        }
        \label{fig:sensitivity-analysis}
    \end{figure}

    To better illustrate this finding, we extracted the slope from the end of each curve (see Fig.~\ref{fig:sensitivity-analysis}(a)) and plotted it against $E$ (Fig.~\ref{fig:sensitivity-analysis}(b)). Repeating this procedure for three different $l_0$ values, we note a sigmoid-like behavior, where the sensitivity switches significantly in a tight $E$-range and remains constant outside. This indicates that there is a critical lowest binding energy $E^*$ required to tune repression using the insulator size; insulator lengths vary about one order of magnitude in cells (200-3000 bp).  The sigmoid curves also show that increasing the enhancer-insulator interaction energy beyond some value does not lead to greater sensitivity.  Moderate energies are enough, which happen to coincide with actual transcription factor binding energies (see Fig.~\ref{supp-fig:energies-from-data}).
    
    To appreciate the magnitude of the switching behavior, we fitted an inverted Hill curve $\sim[1+(E_{1/2}/E)^n]^{-1}$ to the data in Fig. \ref{fig:sensitivity-analysis}(b). The fitting yields quite extreme Hill coefficients in the range $n=10-20$. This indicates ultra-sensitivitve switch-like behavior when $E \approx E_{1/2}$. The fitted Hill coefficients are much higher than observed for typical  TF-operator sites, which are usually smaller than 4~\cite{alon2019introduction}.

    Furthermore, we note that the overall sensitivity decreases with shorter looping times ($E^*$ or $E_{1/2}$ becomes lower with $l_0$).  We envision this reflects cases when the enhancer can explore several sites before finding the promoter, which increases the chance of finding an insulator and, thus, increases insulation.

% =============================================================
\section{Effective resetting theory for enhancer-promoter dynamics} \label{sec:resetting_model}

    \begin{figure*}[!htbp]
        \centering
        \includegraphics[width=0.8\textwidth]{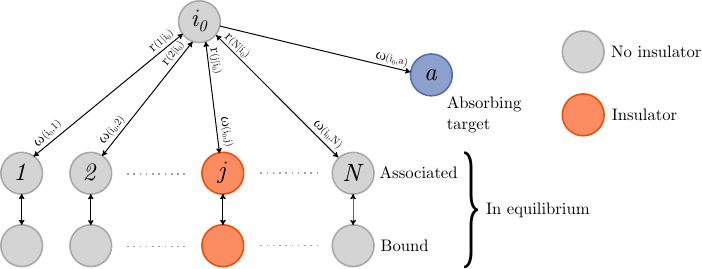}
        \caption{A schematic of the effective model that captures the general model in Sec.~\ref{sec:model}. %
        All rates are shown for an enhancer (or insulator) that starts at $i_0$, which jumps to some node $j$ with a rate $\omega(i_0, j)$ and resets from it with the resetting rate $k(j|i_0)$. Note that the resetting rate is a combination of rates, assuming equilibrium between the associated and bound states (see Sec.~\ref{sec:model} and Supplementary \ref{supp-eq:effective-resetting-rate}.) At some of the sites, an insulator might be present (marked in orange), which affects the resetting rate.
        }
        \label{fig:schematic-effective-model}
    \end{figure*}

    In this section, we derive an analytical theory that allows us to calculate several quantities and better understand our simulation results. For example,  we calculate an exact expression for the first-hitting time density in Laplace transform space $\rho_a(s)$, its first and second moments ($\langle t_a\rangle$ and $\langle t_a^2\rangle$), and the enhancer's position probability density function $P_i(t)$ ($i=1,\ldots, N$). To achieve this, we express the enhancer-insulator dynamics as a target-search problem with resetting in a random energy landscape.

    We envision an enhancer sitting at site $i_0$ looping out to a site $i$ to find a target at $i=a$. When reaching site $i$, the enhancer stays associated for a short while and may either return directly to $i_0$ or become bound at $i$. Regardless of the process, the enhancer eventually returns to $i_0$ like a resetting event, see the schematic in Fig.~\ref{fig:schematic-effective-model}. The difference between the two scenarios is that the residence time is longer if bound, where the enhancer must pass through the "associated" state before it may reset to site $i_0$. Without insulators, the resetting rate is the same across all lattice sites $r$, and this system constitutes a typical resetting problem studied by several authors \cite{evans2011diffusion,chechkin2018random, durang2019first,evans2020stochastic,di2023time}.  In our notation, the master equation for $P_i(t)$ when the resetting rate is constant reads
    \begin{equation}\label{eq:orig-master-eq}
        \begin{split}
        \dv{P_i(t)}{t} = &\sum_{j}\kjump P_j(t) - r P_i(t) \\
                        &+ \sum_{j\neq a} r P_j(t)\delta_{i,i_0} - \rho_a(t)\delta_{i, a},
        \end{split}
    \end{equation}
    where we write the jumping rates as
    \begin{equation}
        \kjump = 
        \begin{cases}
            -\sum_{k\neq i_0} \kloop(k|j),\ & j=i=i_0 \\
            \kloop(i|j),\ &  j=i_0\\
            0,\ &\mathrm{otherwise}.
        \end{cases}
    \end{equation}
    Here,  the diagional element $\omega(i_0,i_0) = -\sum_{k\neq i_0} \kloop(k|i_0)$ represents all outgoing loops from the enhancer's position $i_0$ to anywhere on the lattice, and $\omega(i_0,j) = \kloop(j|i_0)$  is the loop from $i_0$ to $j$. In addition to $\kjump$ and $r$, the master equation includes a sink term $\rho_a(t)\delta_{i, a}$ for the target, where $\rho_a(t)$ is the first-hitting time distribution \cite{chechkin2003first,palyulin2014levy}. Because of this sink, we must compensate with another term proportional to the survival probability $\sum_{j\neq a} P_j(t)$ (3rd term, right-hand side) that ensures there is no resetting if the enhancer has already reached the target.

    Equation \eqref{eq:orig-master-eq} is solvable using standard methods. However, the situation changes when there are insulators. It changes because enhancers and insulators may bind each other, thus creating a non-uniform binding landscape. This manifests as a position-dependent resetting rate $\kreset$, which presents a more complex problem than Eq. \eqref{eq:orig-master-eq} that cannot be solved with standard methods.  

    To find $\kreset$ as a function of the rates and energies outlined in Sec.~\ref{sec:model}, we assume that the "bound" and "associated" states are in equilibrium. This assumption gives (see derivation in \ref{supp-sec:equilibrium-resetting})
    \begin{equation}\label{eq:effective-resetting-rate}
        r(j|i) = \frac{\koff(j|i)}{1 + \frac{\kbind(j|i)}{\kunbind(j|i)}}
                        = \delta\frac{e^{-(\dGl - \Eass)}}{1+e^{\Eass-\Ebound}}.
    \end{equation}
    It is essential to realize that this resetting rate depends critically on the insulators' positions as they determine the binding landscape $\Ebound$ when forming loops with the surrounding chromatin. We show below how we estimate $\Ebound$ assuming the insulators' spatial probability density is in equilibrium. But first, we reformulate the master equation Eq. \eqref{eq:orig-master-eq} with a position-dependent resetting rate in an analytically solvable form.

    To this end, we decompose the resetting rate into two parts
    \begin{equation}
        r(j|i) = r^* + \Delta r(j|i),
    \end{equation}
    where $r^* = {\rm min}_j \left(r(j|i)\right)$ is the smallest resetting rate, and redefine the looping matrix to
    \begin{equation}\label{eq:modified-jump-rate}
        \nu(i,j) =
        \begin{cases}
            \kjump - \Delta r(j|i_0),\ & j=i \\
            \kjump + \Delta r(j|i_0),\ & i=i_0\\
            \kjump \quad &\mathrm{otherwise}.
        \end{cases}
    \end{equation}
    Using this in Eq.~\eqref{eq:orig-master-eq} gives
    \begin{equation}\label{eq:modified-master-equation}
        \begin{split}
            \dv{P_i(t)}{t} = &\sum_{j}\nu(i,j)P_j(t) - r^*P_i(t) \\
                            &+ r^* Q_a(t)\delta_{i,i_0} - \rho_a(t)\delta_{i, a},
        \end{split}
    \end{equation}
    where $Q_a(t) = \sum_{j\neq a} P_j(t)$ is the survival probability. This master equation is now in a standard form and analytically solvable. It represents one of our paper's main results. But before presenting the analytical solution, we outline the basic arguments for including the insulator dynamics in the resetting rate.

\begin{figure}[!htbp]
        \centering
        \includegraphics[width=0.8\columnwidth]{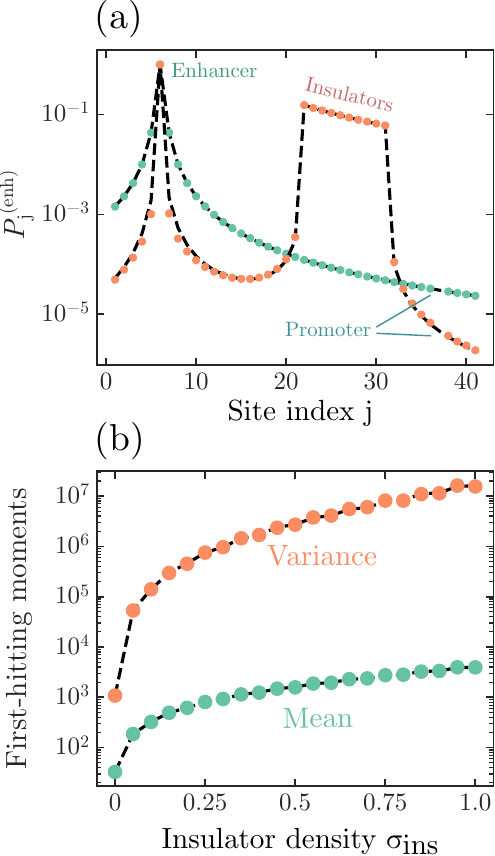}
        \caption{Benchmarking simulations (symbols) and resetting theory (dashed black lines). %
            (a) The probability density function of the enhancer with and without insulators along the lattice.
            (b) The mean and variance of the first-hitting time between enhancer and promoter for varying insulator density.
        }
        \label{fig:ana-vs-sim}
    \end{figure}

    In our simulations, we update the positions of the enhancer and the insulators by drawing loops with rates $\kloop(j|i)$. If the enhancer and one of the insulators happen to end up on the same lattice site, they form a complex with rate $\kbind(j)$.  To include this process in the resetting rate $\kreset$, we make two assumptions. First, we assume that the insulators' probability density function is equilibrated (implying rapid looping dynamics). We calculate the insulator's probability distribution $P^{(\textrm{ins})}_j$ analytically from Eq. \eqref{eq:orig-master-eq} by leaving out the target (see Eq. \eqref{eq:pdf}).  We overlay a few examples of $P^{(\textrm{enh})}_j$ alongside simulated data in Fig. \ref{fig:ana-vs-sim}(a); The agreement is excellent. 

    Second, we assume that the resetting rate $\kreset$ for an enhancer is a weighted combination of the resetting rate with or without an insulator at site $j$. Denoting the rates for these cases as $\kreset_{\rm ins}$ and $\kreset_{\rm no\, ins}$, we obtain
    \begin{equation} \label{eq:binding-landscape}
        \frac 1 \kreset \approx \frac {1- p_\textrm{ins}(j)} {\kreset_{\rm ins}} + \frac{p_\textrm{ins}(j)} {\kreset_{\rm no\, ins}}.
    \end{equation}
    We estimate $p_\textrm{ins}(j)$ as the sum of all the insulator's probability distributions at site $j$, times the probability of a binding event occurring before the insulator leaves:
    \begin{equation}\label{eq:p-ins}
        p_\textrm{ins}(j)= \sum_\textrm{ins} P_j^\textrm{(ins)}(t \to\infty)\frac{\kreset_{\rm ins}}{\kreset_{\rm ins} + r(k|j)},
    \end{equation}
    where $k$ denotes the insulator's starting site (similar to $i_0$ for the enhancer). 
    
    We point out that approximation \eqref{eq:binding-landscape} tends to overestimate the resetting probability. It also changes when insulators have a varying $E^{(\textrm{ins})}$ since it sums over all possible binding energies when, in reality, the enhancer can only be bound to one insulator. 

    In summary, our theoretical model is completely defined by  Eqs.  \eqref{eq:effective-resetting-rate} and \eqref{eq:modified-master-equation}-\eqref{eq:p-ins} that allow us to calculate $\rho_a(t)$ analytically. However,  we note that some of the above assumptions start to fail when the site-specific binding is so strong that the residence times are comparable with typical target-search times (e.g., the free energy barrier $\dGb$ should not be too high (Fig.~\ref{supp-fig:resetting-time-error})).

\subsection{Analytical solution and asymptotic expansion for the first-hitting density}

    In this section, we solve Eq. \eqref{eq:modified-master-equation} analytically.  First, we diagonalize the matrix $\boldsymbol{\nu}$ (with elements in Eq.~\eqref{eq:modified-jump-rate}) into  $\boldsymbol{\nu} = \mathbf  V\mathbf  D\mathbf  V^{-1}$ where each column $\mathbf  V$ is the eigenvector of $\boldsymbol{\nu}$  and $\mathbf  D$ has the eigenvectors $\lambda_1,\ldots, \lambda_N$ (sorted from smallest to largest) along the diagonal (i.e., $D_{ii} = \lambda_i$ and zero otherwise). Next, we introduce the new variable  $q_j(t) = \sum_{i} V_{ji}^{-1}P_i(t)$ and take a Laplace transform ($\mathcal{L}\{f\}(s) = \int_0^{\infty}f(t)e^{-st}\dd t$). This gives
    \begin{equation}\label{eq:laplace-master-equation}
        \begin{split}
        sq_j(s) - V_{ji_0}^{-1} &= \lambda_j q_j(s) - r^*q_j(s) \\
                                    &+ \frac{r^*}{s}V_{ji_0}^{-1} - \rho_a(s)\left[\frac{r^*}{s}V_{ji_0}^{-1} + V_{ja}^{-1} \right],
        \end{split}
    \end{equation}
    where we used that the survival probability $Q_a(t) = 1 - \int_0^t \rho_a(t) dt$ is $Q_a(s) = [1-\rho_a(s)]/s$. As the final steps we obtain a closed form expression for $\rho_a(s)$ by solving Eq.~\eqref{eq:laplace-master-equation} for $q_j(s)$ and use the absorbing boundary condition $\sum_{j} V_{aj}q_j(s) = P_a(s) = 0$. This gives
    \begin{equation}\label{eq:rhoa-laplace}
        \rho_a(s) =
        \frac{(s+r^*)\sum_{j}\frac{A_j}{s+r^*-\lambda_j}}{\sum_{j}\frac{r^*A_j + sB_j}{s+r^*-\lambda_j}},
    \end{equation}
    where $A_j = V_{aj}V_{ji_0}^{-1}$ and $B_j = V_{aj}V_{ja}^{-1}$.  We remind that this analytical solution holds for any enhancer-promoter-insulator configuration (captured in $r(j|i)$ and $\boldsymbol \nu$) under the assumption of fast binding dynamics.
        
    This solution allows us to extract the average hitting time analytically in terms of $\boldsymbol \nu$. By expanding $\rho_a(s)$ for small $s$ (long-time limit) and using 
    \begin{equation}
        \rho_a(s) \simeq 1 -  \langle t_a \rangle s  + \frac{s^2} 2 \langle t_a^2 \rangle -\ldots,
    \end{equation}
    we obtain to first-order in $s$ that~%
    \footnote{%
    Since the edge case $r^* = 0$ is unsolvable if $\lambda_j = 0$, we remove $\lambda_1 = 0$ from the sum.
    }
    \begin{equation}\label{eq:first_moment}
        \langle t_a \rangle = \frac{\sum_{j \neq 1}\frac{B_j-A_j}{r^*-\lambda_j}}{A_n + r^*\sum_{j \neq 1}\frac{A_j}{r^*-\lambda_j}}.
    \end{equation}
    A similar expansion focusing on the second-order terms yields the second moment
    \begin{equation}\label{eq:second-moment}
        \begin{split}
            \langle t_a^2 \rangle =
            &\frac{2}{\left(A_1 + \sum_{j\neq 1} \frac{r^*A_j}{r^* - \lambda_j}\right)^2} \cross \\
            \Bigg[
                &\left(\sum_{j\neq 1}\left(-\frac{r^*A_j}{(r^* - \lambda_j)^2} + \frac{B_j}{r^* - \lambda_j}\right)\right) \cross \\
                &\left(\sum_{j\neq 1} \frac{B_j - A_j}{r^* - \lambda_j}\right) + \\
                &\left(A_1 + \sum_{j\neq 1} \frac{r^*A_j}{r^* - \lambda_j}\right)
                \left(\sum_{j\neq 1} \frac{B_j - A_j}{(r^* - \lambda_j)^2}\right)
            \Bigg].
        \end{split}
    \end{equation}
    In supplementary (see \ref{supp-sec:further-expansion}), we provide an explicit formula for the variance $\sigma^2 = \langle t_a^2 \rangle  - (\langle t_a \rangle)^2$. 

    To test the theory against simulations, we compare the variance and mean in Fig. \ref{fig:ana-vs-sim}(b). The simulations (symbols) and the theory (black dashed lines) show excellent agreement.

    As a final result, we show that the analytical result for the probability density $P_i(t)$ is valid for both enhancers and insulators. If ignoring the absorbing target (thus setting $Q_a(s) = 1$), Eq. \eqref{eq:laplace-master-equation} becomes
    \begin{equation}\label{eq:pdf}
        P_i(t) = \sum_{j} V_{ij}V_{ji_0}^{-1} \frac{r^* - \lambda_j e^{(\lambda_j - r^*)t}}{r^* - \lambda_j},
    \end{equation}
    where the initial condition here is $P_i(0) = \delta_{i, i_0}$. In the steady-state, this equation simplifies to 
    \begin{equation}\label{eq:pdf-equilib}
        P_i(t\to \infty) =  V_{i1}V_{1i_0}^{-1} + \sum_{j\neq 1}\frac{r^*}{r^* - \lambda_j}V_{ij}V_{ji_0}^{-1},
    \end{equation}
    which we used to estimate the insulators' positions in the resetting rate $\kreset$. We show the complete derivation in the supplementary material (Sec. \ref{supp-sec:prob-dist}).

% ==========================================================================
% Discussion
\section{Discussion and Conclusion}

Cells use a complex web of enhancer-insulator interactions to support gene regulatory networks and orchestrate signaling cascades during development. Insulator elements are critical to prevent unintended gene activation and are relatively simple from a genetic point of view---removing them from DNA or abolishing the associated transcription factors causes gene activation. Yet, it remains unclear how this happens because of recent conflicting empirical observations \cite{kahn2023topological}. This paper presents a new mechanistic model where insulators bind weakly to surrounding chromatin and the enhancer rather than insulators attaching themselves, which is a common assumption.

This assumption largely derives from so-called insulator bypass \cite{muravyova2001loss, cai2001effects, kyrchanova2008orientation}. This phenomenon refers to a family of genetic experiments aiming to neutralize insulator-enhancer blocking by genetically inserting a piece of foreign DNA. Known as transgenic constructs, these DNA pieces contain yet another insulator (specifically,  a short DNA sequence that attracts insulator-binding factors), and several studies show that these constructs help remove the blocking when sandwiched between the insulator-enhancer pair.  The governing explanation for these observations is that insulators pair up and form a loop. This theory has yet further support from measurements showing that some insulator-binding factors can bind each other (e.g., \cite{kyrchanova2008orientation}). However, most transgenic experiments study short-ranged interactions, typically less than 5 kb. While insulator pairing could be the primary insulation mechanism at short distances, it is doubtfully so over long distances. This is the principal observation in \cite{kahn2023topological} using Hi-C data, having 5kb as the lower resolution limit. They could not detect notable insulator-insulator interactions across thousands of pairs.

These observations form the starting point of this work. We aimed to establish a new biophysical model that did not rely on specific insulator-insulator interactions. Instead, it rests on generic but weak insulator-chromatin interactions. From a few assumptions, we calibrated the model to a few independent empirical datasets and derived analytical results that fit empirical observations. 

One assumption that likely represents an oversimplification is that the insulators have equal binding strengths to enhancers and the surrounding chromatin. We also limit this study to simple enhancer-insulator-promoter configurations. But in reality, gene clusters have more complex arrangements, including many enhancers, insulators, and promoters, all having heterogeneous binding strengths. It would be instructive to investigate how hitting frequencies between select enhancer-promoter pairs respond to changes in these variables. For example, will one strong insulator do the same job as a few weak ones? We leave these questions for future work.

To close, mammalian genomes harbor $\sim 20,000$ genes regulated by $\sim 900,000$ enhancer-like elements interspersed with $\sim 30,000$ CTCF-bound sites, many of which act as insulators \cite{encode2020expanded}. These elements represent critical components of gene regulatory networks that also seem to shape DNA's spatial organization by forming complex networks of 3D interactions and semi-hierarchical 3D communities  \cite{bernenko2023mapping} (e.g., Topologically Associated Domains and A/B compartments). Thus, unveiling the mechanisms of insulation is a critical step to understanding the causal mechanisms of the structure-function relationship of interphase chromosomes.

% ==========================================================================
% Acknowledgements
\acknowledgments%
We acknowledge financial support from the Swedish Research Council (grant no.  2017-03848  and no. 2021-04080). %
RM acknowledges funding from the German Science Foundation (DFG, grant ME 1535/12-1).
We want to thank Dr.  Yuri Schwartz and Dr. Tatiana Kahn (Umeå University) for providing experimental data in Fig. 1(c) and valuable discussions.
The computations were enabled by resources provided by the National Academic Infrastructure for Supercomputing in Sweden (NAISS) and the Swedish National Infrastructure for Computing (SNIC) at High-Performance Computing Center North (HPC2N), partially funded by the Swedish Research Council through grant agreements no. 2022-06725 and no. 2018-05973.

L.L. and L.H. devised the study. L.H. performed the simulations, analytical solutions and visualizations. All authors contributed in writing the manuscript.

% ==========================================================================
% Appendix
%\appendix%
%\subfile{sections/appendix.tex}

% ==========================================================================
% References
\bibliography{refs}

\end{document}